\documentclass[letterpaper,twocolumn,prb,longbibliography]{revtex4-2}
\AtBeginDocument{
\heavyrulewidth=.08em
\lightrulewidth=.05em
\cmidrulewidth=.03em
\belowrulesep=.65ex
\belowbottomsep=0pt
\aboverulesep=.4ex
\abovetopsep=0pt
\cmidrulesep=\doublerulesep
\cmidrulekern=.5em
\defaultaddspace=.5em
}

\usepackage{float}
\usepackage{comment}
\usepackage{booktabs}
\usepackage[percent]{overpic}
\usepackage{microtype}
\usepackage{array}
\usepackage{amsmath}
\usepackage{amsthm}
\usepackage{amssymb}
\usepackage{physics}
\usepackage{graphicx}

\theoremstyle{remark}

\theoremstyle{definition}

\usepackage{bbold}
\numberwithin{thm}{section}
\usepackage{hyperref}
\usepackage[usenames,dvipsnames]{xcolor}
\usepackage{tikz}
\usetikzlibrary{arrows.meta, automata, positioning, quotes, shapes}

\newcommand{\TikzFigOne}{
\begin{tikzpicture}[baseline=(current bounding box.center), scale=1.5]
\draw [black, ultra thick] (0,0) -- (0.5,0.5);
\draw [black, ultra thick] (0,0) -- (-0.5,0.5);
\draw [black, ultra thick] (-0.5,0.5) -- (0,1);
\draw [black, ultra thick] (0.5,0.5) -- (0,1);

\draw [gray,dashed,thin] (-0.5,0.5) -- (0.5,0.5);

\fill[white] (-0.05, -0.05) rectangle (0.05,0.05);
\fill[white] (-0.55, 0.45) rectangle (-0.45,0.55);
\fill[white] (0.45, 0.45) rectangle (0.55,0.55);
\fill[white] (-0.05, 0.95) rectangle (0.05,1.05);

\foreach \i in {0,...,0} {
    \foreach \j in {0,...,1} {
	    \draw[thick] (\i-0.05,\j-0.05) rectangle ++(0.1,0.1);
    }
}

\foreach \i in {-1,...,0} {
    \foreach \j in {0,...,0} {
	    \draw[thick] (\i+0.45,\j+0.45) rectangle ++(0.1,0.1);
    }
}

\fill[gray] (0,0.5) circle (0.05);

\node at (0.0,-0.2) [anchor=center, text=black, font=\small] {$\tilde{\sigma}^x$};
\node at (0.0,1.2) [anchor=center, text=black, font=\small] {$\tilde{\sigma}^x$};
\node at (-0.75,0.5) [anchor=center, text=black, font=\small] {$\tilde{\sigma}^x$};
\node at (0.75,0.5) [anchor=center, text=black, font=\small] {$\tilde{\sigma}^x$};
\end{tikzpicture}
}

\newcommand{\TikzFigTwo}{
\begin{tikzpicture}[baseline=(current bounding box.center), scale=1.5]

\draw [gray,dashed,thin] (0,0) -- (0.5,0.5);
\draw [gray,dashed,thin] (0,0) -- (-0.5,0.5);
\draw [gray,dashed,thin] (-0.5,0.5) -- (0,1);
\draw [gray,dashed,thin] (0.5,0.5) -- (0,1);

\draw [black,ultra thick] (-0.5,0.5) -- (0.5,0.5);

\fill[white] (-0.05, -0.05) rectangle (0.05,0.05);
\fill[white] (-0.55, 0.45) rectangle (-0.45,0.55);
\fill[white] (0.45, 0.45) rectangle (0.55,0.55);
\fill[white] (-0.05, 0.95) rectangle (0.05,1.05);

\foreach \i in {0,...,0} {
    \foreach \j in {0,...,1} {
	    \draw[gray,thin] (\i-0.05,\j-0.05) rectangle ++(0.1,0.1);
    }
}

\foreach \i in {-1,...,0} {
    \foreach \j in {0,...,0} {
	    \draw[gray,thin] (\i+0.45,\j+0.45) rectangle ++(0.1,0.1);
    }
}

\fill[black] (0,0.5) circle (0.07);
\node at (0,0.65) [anchor=center, text=black, font=\small] {$\sigma^y$};
\end{tikzpicture}
}

\newcommand{\TikzFigThree}[1]{
\begin{tikzpicture}[baseline=(current bounding box.center), scale=1, rotate=#1]
\foreach \i in {0,...,0} {
    \foreach \j in {0,...,1} {
	    \draw (\i-0.05,\j-0.05) rectangle ++(0.1,0.1);
    }
}

\foreach \i in {-1,...,0} {
    \foreach \j in {0,...,0} {
	    \draw (\i+0.45,\j+0.45) rectangle ++(0.1,0.1);
    }
}

\draw[gray,thin] (1-0.05,1-0.05) rectangle ++(0.1,0.1);
\draw[gray,thin] (0.5-0.05,1.5-0.05) rectangle ++(0.1,0.1);
\draw[gray,thin] (1.5-0.05,0.5-0.05) rectangle ++(0.1,0.1);
\draw[gray,thin] (1-0.05,-0.05) rectangle ++(0.1,0.1);
\draw[gray,thin] (0.5-0.05,-0.5-0.05) rectangle ++(0.1,0.1);

\draw [black, ultra thick] (0,0) -- (0.5,0.5);
\draw [black, ultra thick] (0,0) -- (-0.5,0.5);
\draw [black, ultra thick] (-0.5,0.5) -- (0,1);
\draw [black, ultra thick] (0.5,0.5) -- (0,1);

\draw [gray,dashed,thin] (-0.5,0.5) -- (1.5,0.5);
\draw [gray,dashed,thin] (0.5,-0.5) -- (0.5,1.5);

\fill[white] (-0.05, -0.05) rectangle (0.05,0.05);
\fill[white] (-0.55, 0.45) rectangle (-0.45,0.55);
\fill[white] (0.45, 0.45) rectangle (0.55,0.55);
\fill[white] (-0.05, 0.95) rectangle (0.05,1.05);

\fill[gray] (0,0.5) circle (0.05);
\fill[gray] (1,0.5) circle (0.05);
\fill[gray] (0.5,0) circle (0.05);
\fill[gray] (0.5,1) circle (0.05);

\end{tikzpicture}
}

\newcommand{\TikzFigYcheck}{
\begin{tikzpicture}[baseline=(current bounding box.center), scale=1]
\foreach \i in {0,...,0} {
    \foreach \j in {0,...,1} {
	    \draw[gray] (\i-0.05,\j-0.05) rectangle ++(0.1,0.1);
    }
}

\foreach \i in {-1,...,0} {
    \foreach \j in {0,...,0} {
	    \draw[gray] (\i+0.45,\j+0.45) rectangle ++(0.1,0.1);
    }
}

\draw[gray,thin] (1-0.05,1-0.05) rectangle ++(0.1,0.1);
\draw[gray,thin] (0.5-0.05,1.5-0.05) rectangle ++(0.1,0.1);
\draw[gray,thin] (1.5-0.05,0.5-0.05) rectangle ++(0.1,0.1);
\draw[gray,thin] (1-0.05,-0.05) rectangle ++(0.1,0.1);
\draw[gray,thin] (0.5-0.05,-0.5-0.05) rectangle ++(0.1,0.1);

\draw [black, ultra thick] (-0.5,0.5) -- (1.5,0.5);
\draw [black, ultra thick] (0.5,-0.5) -- (0.5,1.5);

\fill[white] (-0.55, 0.45) rectangle (-0.45,0.55);
\fill[white] (1.45, 0.45) rectangle (1.55,0.55);
\fill[white] (0.45, -0.55) rectangle (0.55,-0.45);
\fill[white] (0.45, 1.45) rectangle (0.55,1.55);

\fill[black] (0,0.5) circle (0.05);
\fill[black] (1,0.5) circle (0.05);
\fill[black] (0.5,0) circle (0.05);
\fill[black] (0.5,1) circle (0.05);
\end{tikzpicture}
}

\newcommand{\TikzFigXXXX}{
\begin{tikzpicture}[baseline=(current bounding box.center), scale=1]
\draw[gray,thin] (-0.5-0.05,0.5-0.05) rectangle ++(0.1,0.1);
\draw[gray,thin] (0.5-0.05,1.5-0.05) rectangle ++(0.1,0.1);
\draw[gray,thin] (1.5-0.05,0.5-0.05) rectangle ++(0.1,0.1);
\draw[gray,thin] (0.5-0.05,-0.5-0.05) rectangle ++(0.1,0.1);
\draw[gray,thin] (0.5-0.05,0.5-0.05) rectangle ++(0.1,0.1);

\draw [black, ultra thick] (-0.5,0.5) -- (0.5,1.5);
\draw [black, ultra thick] (1.5,0.5) -- (0.5,1.5);
\draw [black, ultra thick] (1.5,0.5) -- (0.5,-0.5);
\draw [black, ultra thick] (-0.5,0.5) -- (0.5,-0.5);

\draw [gray,dashed,thin] (-0.5,0.5) -- (1.5,0.5);
\draw [gray,dashed,thin] (0.5,-0.5) -- (0.5,1.5);

\fill[white] (-0.55, 0.45) rectangle (-0.45,0.55);
\fill[white] (0.45, 1.45) rectangle (0.55,1.55);
\fill[white] (0.45, -0.55) rectangle (0.55,-0.45);
\fill[white] (1.45, 0.45) rectangle (1.55,0.55);

\fill[gray] (0,0.5) circle (0.05);
\fill[gray] (1,0.5) circle (0.05);
\fill[gray] (0.5,0) circle (0.05);
\fill[gray] (0.5,1) circle (0.05);
\end{tikzpicture}
}

\newcommand{\TikzAppa}{
\begin{tikzpicture}[baseline=(current bounding box.center), scale=1]
\draw [very thin, black] (0,0) grid (2,3);
\draw[black,ultra thick] (2, 0) -- (2, 3);

\foreach \i in {0,1} {
    \foreach \j in {0,1,...,2} {
        \pgfmathtruncatemacro{\checker}{mod(\i+\j,2)}
        \ifnum\checker=1
            \fill[fill=gray, opacity=0.5] (\i,\j) -- (\i+1,\j) -- (\i+1,\j+1) -- (\i,\j+1) -- cycle;
        \fi
    }
}

\foreach \i in {0,...,2} {
    \foreach \j in {0,...,3} {
        \fill (\i,\j) circle (0.05);
    }
}

\fill[green] (1.4, 2.4) rectangle (1.6,2.6);
\fill[green] (1.4, 0.4) rectangle (1.6,0.6);

\node at (1.5,1.5) [anchor=center, text=green, font=\small, rotate=0] {$S^{(2)}$};
\end{tikzpicture}
}

\newcommand{\TikzAppb}{
\begin{tikzpicture}[baseline=(current bounding box.center), scale=1]
\draw [very thin, black] (0,0) grid (2,3);
\draw[black,ultra thick] (2, 0) -- (2, 3);

\foreach \i in {0,1} {
    \foreach \j in {0,1,...,2} {
        \pgfmathtruncatemacro{\checker}{mod(\i+\j,2)}
        \ifnum\checker=1
            \fill[fill=gray, opacity=0.5] (\i,\j) -- (\i+1,\j) -- (\i+1,\j+1) -- (\i,\j+1) -- cycle;
        \fi
    }
}

\foreach \i in {0,...,2} {
    \foreach \j in {0,...,3} {
        \fill (\i,\j) circle (0.05);
    }
}

\draw[green,ultra thick] (2, 0) -- (2, 2);

\draw[green] (1.4, 2.4) rectangle (1.6,2.6);
\fill[green] (2,0) circle (0.1);
\fill[green] (2,1) circle (0.1);
\fill[green] (2,2) circle (0.1);

\draw[green,dashed] (1.5, 2.5) -- (1.5, 0);
\draw[green,dashed] (1.5, 2.5) -- (2, 2.5);

\end{tikzpicture}
}

\newcommand{\TikzAppc}{
\begin{tikzpicture}[baseline=(current bounding box.center), scale=1]
\draw [very thin, black] (0,0) grid (2,3);
\draw[black,ultra thick] (2, 0) -- (2, 3);

\foreach \i in {0,1} {
    \foreach \j in {0,1,...,2} {
        \pgfmathtruncatemacro{\checker}{mod(\i+\j,2)}
        \ifnum\checker=1
            \fill[fill=gray, opacity=0.5] (\i,\j) -- (\i+1,\j) -- (\i+1,\j+1) -- (\i,\j+1) -- cycle;
        \fi
    }
}

\foreach \i in {0,...,2} {
    \foreach \j in {0,...,3} {
        \fill (\i,\j) circle (0.05);
    }
}

\draw[green] (1.4,0.4) rectangle (1.6,0.6);
\fill[green] (2,0) circle (0.1);

\draw[green,dashed] (1.5, 0.5) -- (1.5, 0);
\draw[green,dashed] (1.5, 0.5) -- (2, 0.5);
\end{tikzpicture}
}

\newcommand{\TikzAppd}{
\begin{tikzpicture}[baseline=(current bounding box.center), scale=1]
\draw [very thin, black] (0,0) grid (2,3);
\draw[black,ultra thick] (2, 0) -- (2, 3);

\foreach \i in {0,1} {
    \foreach \j in {0,1,...,2} {
        \pgfmathtruncatemacro{\checker}{mod(\i+\j,2)}
        \ifnum\checker=1
            \fill[fill=gray, opacity=0.5] (\i,\j) -- (\i+1,\j) -- (\i+1,\j+1) -- (\i,\j+1) -- cycle;
        \fi
    }
}

\foreach \i in {0,...,2} {
    \foreach \j in {0,...,3} {
        \fill (\i,\j) circle (0.05);
    }
}

\fill[green] (2,1) circle (0.1);
\fill[green] (2,2) circle (0.1);

\draw[green,ultra thick] (2, 1) -- (2, 2);

\node at (1.6,1.5) [anchor=center, text=green, font=\small, rotate=90] {$\sigma_i^y \sigma_{i+1}^y$};
\end{tikzpicture}
}

\newcommand{\TikzAppA}{
\begin{tikzpicture}[baseline=(current bounding box.center), scale=1]
\draw [very thin, black] (0,0) grid (2,3);
\draw[black,ultra thick] (0, 0) -- (0, 3);

\foreach \i in {0,1} {
    \foreach \j in {0,1,...,2} {
        \pgfmathtruncatemacro{\checker}{mod(\i+\j,2)}
        \ifnum\checker=1
            \fill[fill=gray, opacity=0.5] (\i,\j) -- (\i+1,\j) -- (\i+1,\j+1) -- (\i,\j+1) -- cycle;
        \fi
    }
}

\foreach \i in {0,...,2} {
    \foreach \j in {0,...,3} {
        \fill (\i,\j) circle (0.05);
    }
}

\fill[green] (1.4,0.4) rectangle (1.6,0.6);
\fill[red] (0.4,1.4) rectangle (0.6,1.6);
\fill[green] (1.4,2.4) rectangle (1.6,2.6);

\node at (1.5,1.5) [anchor=center, text=teal, font=\small, rotate=0] {$S^{(1)}$};
\end{tikzpicture}
}

\newcommand{\TikzAppB}{
\begin{tikzpicture}[baseline=(current bounding box.center), scale=1]
\draw [very thin, black] (0,0) grid (2,3);
\draw[black,ultra thick] (0, 0) -- (0, 3);

\foreach \i in {0,1} {
    \foreach \j in {0,1,...,2} {
        \pgfmathtruncatemacro{\checker}{mod(\i+\j,2)}
        \ifnum\checker=1
            \fill[fill=gray, opacity=0.5] (\i,\j) -- (\i+1,\j) -- (\i+1,\j+1) -- (\i,\j+1) -- cycle;
        \fi
    }
}

\foreach \i in {0,...,2} {
    \foreach \j in {0,...,3} {
        \fill (\i,\j) circle (0.05);
    }
}

\draw[green,dashed] (1.5, 0.5) -- (0, 0.5);
\draw[green,dashed] (1.5, 0.5) -- (1.5, 0.0);

solid lines
\draw[green,ultra thick] (1, 0) -- (0, 0.0);

\draw[green] (1.4,0.4) rectangle (1.6,0.6);
\fill[green] (0,0) circle (0.1);
\fill[green] (1,0) circle (0.1);
\end{tikzpicture}
}

\newcommand{\TikzAppC}{
\begin{tikzpicture}[baseline=(current bounding box.center), scale=1]
\draw [very thin, black] (0,0) grid (2,3);
\draw[black,ultra thick] (0, 0) -- (0, 3);

\foreach \i in {0,1} {
    \foreach \j in {0,1,...,2} {
        \pgfmathtruncatemacro{\checker}{mod(\i+\j,2)}
        \ifnum\checker=1
            \fill[fill=gray, opacity=0.5] (\i,\j) -- (\i+1,\j) -- (\i+1,\j+1) -- (\i,\j+1) -- cycle;
        \fi
    }
}
\foreach \i in {0,...,2} {
    \foreach \j in {0,...,3} {
        \fill (\i,\j) circle (0.05);
    }
}

\draw[green,dashed] (1.5, 2.5) -- (0, 2.5);
\draw[green,dashed] (1.5, 2.5) -- (1.5, 0.0);

\draw[green,ultra thick] (1, 0) -- (0, 0);
\draw[green,ultra thick] (1, 1) -- (0, 1);
\draw[green,ultra thick] (1, 2) -- (0, 2);
\draw[green,ultra thick] (0, 0) -- (0, 2);
\draw[green,ultra thick] (1, 0) -- (1, 2);

\draw[green] (1.4,2.4) rectangle (1.6,2.6);
\fill[green] (0,0) circle (0.1);
\fill[green] (1,0) circle (0.1);
\fill[green] (1,1) circle (0.1);
\fill[green] (1,2) circle (0.1);
\fill[green] (0,1) circle (0.1);
\fill[green] (0,2) circle (0.1);
\end{tikzpicture}
}

\newcommand{\TikzAppD}{
\begin{tikzpicture}[baseline=(current bounding box.center), scale=1]
\draw [very thin, black] (0,0) grid (2,3);
\draw[black,ultra thick] (0, 0) -- (0, 3);

\foreach \i in {0,1} {
    \foreach \j in {0,1,...,2} {
        \pgfmathtruncatemacro{\checker}{mod(\i+\j,2)}
        \ifnum\checker=1
            \fill[fill=gray, opacity=0.5] (\i,\j) -- (\i+1,\j) -- (\i+1,\j+1) -- (\i,\j+1) -- cycle;
        \fi
    }
}

\foreach \i in {0,...,2} {
    \foreach \j in {0,...,3} {
        \fill (\i,\j) circle (0.05);
    }
}

\draw[red,ultra thick] (1, 1) -- (0, 2);
\draw[red,ultra thick] (0, 1) -- (1, 2);

\draw[red] (0.4,1.4) rectangle (0.6,1.6);
\fill[red] (0,1) circle (0.1);
\fill[red] (1,1) circle (0.1);
\fill[red] (1,2) circle (0.1);
\fill[red] (0,2) circle (0.1);
\end{tikzpicture}
}

\newcommand{\TikzAppE}{
\begin{tikzpicture}[baseline=(current bounding box.center), scale=1]
\draw [very thin, black] (0,0) grid (2,3);
\draw[black,ultra thick] (0, 0) -- (0, 3);

\foreach \i in {0,1} {
    \foreach \j in {0,1,...,2} {
        \pgfmathtruncatemacro{\checker}{mod(\i+\j,2)}
        \ifnum\checker=1
            \fill[fill=gray, opacity=0.5] (\i,\j) -- (\i+1,\j) -- (\i+1,\j+1) -- (\i,\j+1) -- cycle;
        \fi
    }
}

\foreach \i in {0,...,2} {
    \foreach \j in {0,...,3} {
        \fill (\i,\j) circle (0.05);
    }
}

\draw[green,ultra thick] (1, 1) -- (1, 2);
\draw[green,ultra thick] (0, 1) -- (0, 2);
\draw[green,ultra thick] (0, 1) -- (1, 1);
\draw[green,ultra thick] (0, 2) -- (1, 2);

\fill[green] (1,1) circle (0.1);
\fill[green] (1,2) circle (0.1);
\fill[green] (0,1) circle (0.1);
\fill[green] (0,2) circle (0.1);
\end{tikzpicture}
}

\newcommand{\TikzAppF}{
\begin{tikzpicture}[baseline=(current bounding box.center), scale=1]
\draw [very thin, black] (0,0) grid (2,3);
\draw[black,ultra thick] (0, 0) -- (0, 3);

\foreach \i in {0,1} {
    \foreach \j in {0,1,...,2} {
        \pgfmathtruncatemacro{\checker}{mod(\i+\j,2)}
        \ifnum\checker=1
            \fill[fill=gray, opacity=0.5] (\i,\j) -- (\i+1,\j) -- (\i+1,\j+1) -- (\i,\j+1) -- cycle;
        \fi
    }
}

\foreach \i in {0,...,2} {
    \foreach \j in {0,...,3} {
        \fill (\i,\j) circle (0.05);
    }
}

\draw[blue,ultra thick] (1, 1) -- (0, 2);
\draw[blue,ultra thick] (0, 1) -- (1, 2);

\fill[blue] (0,1) circle (0.1);
\fill[blue] (1,1) circle (0.1);
\fill[blue] (1,2) circle (0.1);
\fill[blue] (0,2) circle (0.1);

\node at (1.2,1.5) [anchor=center, text=blue, font=\small, rotate=90] {$B_s$};
\end{tikzpicture}
}

\usepackage{hyperref} 
\hypersetup{colorlinks=true,citecolor=Blue,linkcolor=Blue,urlcolor=Blue}
\newcommand{\rsquare}{\vcenter{\hbox{\rotatebox{45}{\scalebox{0.7}{$\square$}}}}}

\usepackage[normalem]{ulem}

\begin{document}

\title{Hidden subsystem symmetry protected states in competing topological orders 
}
\author{Shi Feng}
\affiliation{Department of Physics, The Ohio State University, Columbus, Ohio 43210, USA}

\begin{abstract}
We reveal the connection between two-dimensional subsystem symmetry-protected topological (SSPT) states and two-dimensional topological orders via a self-dual frustrated toric code model. This model, an enrichment of the toric code (TC) with its dual interactions, can be mapped to a model defined on the dual lattice with subsystem symmetries and subextensive ground state degeneracy. The map connects exactly the frustrated TC to two copies of the topological plaquette Ising model (TPIM), as a strong SSPT model with linear subsystem symmetries. The membrane order parameter of the TPIM is exactly mapped to dual TC stabilizers as the order parameter of the frustrated TC model, SSPT gapless edge states of the TPIM are mapped to zero-energy dangling operators under open boundaries, and the transition from the SSPT-ordered TPIM to the trivial paramagnetic phase is mapped to the transition between two distinct topological orders. We also demonstrate that this mapping can be used to elucidate the structure of other SSPT models, reflecting the subtle linkage between SSPT order and topological order in two dimensions.

\end{abstract}
\date{\today}
\maketitle
\section{Introduction}
The exploration of topological phases of matter has expanded our understanding of distinct states of matter beyond those conventionally defined by spontaneously broken symmetry. Three primary categories of topological phases have emerged:
(1) Phases protected by an unbroken global symmetry, i.e. symmetry protected topological (SPT) phases exemplified by topological band insulators \cite{Kane2005} and integer spin chains \cite{Affleck}.
(2) Topologically ordered phases \cite{wen1990,wen2017colloquium}, as found in matters like quantum spin liquids \cite{Wen_PRB_2002,kitaev2006anyons,ZhouRMP}, Kitaev's toric code \cite{kitaev2003fault,Alexei2002}, and fractional quantum Hall systems \cite{Wen1990f}, known for their inherent or emergent local gauge fields and braiding statistics.
(3) The most recent and compelling addition, the Subsystem Symmetry Protected Topological (SSPT) phases \cite{Yizhi20180,You2018}, which lies at an intermediate position between the first two categories, featuring a sub-global symmetry and non-local order parameter. Just as SPT states are known for boundary modes that are protected by global symmetries, SSPT states are also protected in a similar fashion, but by their subextensive subsystem symmetries. Such phases introduce new forms of matter characterized by subextensive topological ground-state degeneracy (GSD), quasiparticle excitations with fractonic mobility \cite{Sagar2016,Yan2019a,Yan2019b,Xie2019,Shirley2019,Slagle2021}.

\begin{figure}[t]
\centering
\includegraphics[width=0.95\linewidth]{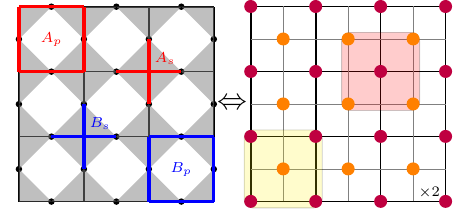}
\caption{ Mapping between the self-dual frustrated TC model (left) and two copies of the TPIM (right). (Left): TC square lattice $\Lambda$ in solid black lines, where qubits live on links; its equivalence in terms of checker-board lattice in Wen's plaquette representation \cite{Wen2003} -- star operator corresponds to shaded square while plaquette operator to white square; and TC stabilizers $A_s$ (red star), $B_p$ (blue plaquette), and dual TC stabilizers $A_p$ (red plaquette) and $B_s$ (blue star).  (Right): The TPIM lattice. Two sublattices are denoted by black and gray lines, with qubits labeled in orange ($\sigma$) and purple $\tau$. The two five-qubit interactions: $\tau_{\rm center}^z\prod_{i\in \rm corners} \sigma_{i}^x$ and $\sigma_{\rm center}^z\prod_{i\in \rm corners} \tau_{i}^x$, are circled in yellow and red.}
\label{fig:sum}
\end{figure}

The primary objective of this paper is to elucidate an intricate relationship between 2D SSPT states and 2D topological order. We concentrate on the “strong SSPT” states \cite{Yizhi20180,You2018}, distinct from the “weak SSPT” states in that they cannot be simply regarded as decoupled or weakly coupled SPT states in lower dimensions.
While several related scenarios are well-understood, such as the connection between 2D spontaneous symmetry breaking (SSB) models, 2D global SPT order and its dual 2D topological order \cite{Chen2007,Wen2005,Gu2012}, and between 3D SSPT orders and 3D fracton orders \cite{Nathanan2020,Shirley2020}, the relationship between 2D SSPT orders and 2D topological orders remains relatively less discussed.
In this work, in loose analogy to the hidden symmetry-breaking picture observed in an SPT phase after a non-local transformation \cite{Kennedy1992a,Kennedy1992b,Pollmann2012}, we demonstrate the existence of hidden competing topological orders within 2D SSPT phases, or conversely, the hidden 2D SSPT phases within competing 2D topological orders.
We note that previous studies have established a connection between the triangular cluster-state model and its dual topological order in the context of defect homology \cite{Nathanan2020}, we here take a different approach: by using an explicit non-local map to relate the simplest topological order, the toric code (TC), to the simplest strong SSPT model, the topological plaquette Ising model (TPIM), in the dual lattice [Fig~.\ref{fig:sum}].

Let us consider the following self-dual frustrated TC Hamiltonian (FTC), where we superimpose the TC interactions with its dual interactions 
\begin{equation}\label{eq:FTC}
        H_{\rm FTC} = -\left[\sum_s A_s + \sum_p B_p \right] - \alpha \left[\sum_p A_p + \sum_s B_s\right]
\end{equation}
The first two terms, $A_s  =\prod_{i\in s} \sigma_i^x,~ B_p = \prod_{i\in p} \sigma_i^z$, are the $X$ and $Z$ parity checks in the canonical $H_{\rm TC}$ of TC, while the dual interactions, in the later two terms,
\begin{equation} \label{eq:dualtc}
    A_p =\prod_{i\in p} \sigma_i^x,~ B_s = \prod_{i\in s} \sigma_i^z
\end{equation}
are added as perturbations parameterized by $\alpha$. It depicts the competition between two mutually frustrating TCs and enjoys a duality $\alpha \leftrightarrow \alpha^{-1}$. 
At transition $\alpha = 1$ both vertices and plaquettes play the same role. 
The perturbation in Eq.~\ref{eq:FTC} is not disjoint from the original 2D TC and it does not commute with $H_{\rm TC}$, e.g. the dual operator $A_p$ and its neighboring plaquette $B_{p'}$ share a single link. 
We note that recently there has been investigation on the competition between topological orders in 2D TC and 3D X-cube model \cite{Vidal2022}, where emergent phases with subextensive symmetries are reported. 
Here we show that the competition between topological orders in 2D is already interesting. 
Note that Kitaev's original TC has ${\rm GSD} = 4$ under periodic boundary, however, it acquires infinite GSD under open boundaries due to gapless edge modes \cite{Wen2008}.   
We will show that the two competing TCs in Eq.~\ref{eq:FTC} under open boundary can be mapped to a model with a subextensively infinite GSD, where the dangling edge operators of an open-boundary TC lattice is mapped to the fractionalized edge symmetries of the TPIM.


\begin{figure}[t]
\centering
\includegraphics[width=\linewidth]{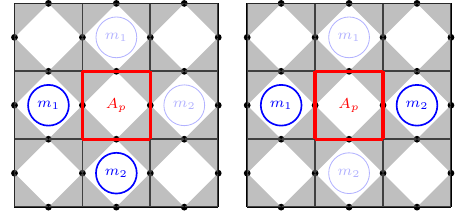}
\caption{The anyon mobility analysis in leading order perturbation theory. Left: A local perturbation by the dual TC operator $A_p$ (in red) moves the two anyon bound state $m_1\times m_2$ aligned in diagonal direction (deep blue) to its adjacent position (faint blue) along the off-diagonal direction, resulting in a bound-state dispersion along the one-dimensional direction. Right: The same perturbation transitions the horizontally aligned two-anyon bound state (deep blue) to a vertically aligned state (faint blue), and conversely. This results in a stable, immobile bound state.  }
\label{fig:anyon}
\end{figure}

\section{Excitation and order parameter}
In this section we discuss the quasi-particle excitations of Eq.~\ref{eq:FTC}, where anyon-bound states exhibit reduced dimensionality or fractonic mobility, reflecting hidden subsystem symmetries in the dual lattice; and the order parameter of Eq.~\ref{eq:FTC} which distinguish two competing TC topological order. 
\subsection{Fractonic anyons}
A common signature of the emergence of subsystem symmetries is the effective dimensional reduction, i.e. the fractonic nature, of excitations \cite{Yan2019a,Sobrinho2020,Hermele2022}. Here we start from the mobility of anyons in the leading order perturbation theory as a rudimentary indicator, showing that the competition between the two $Z_2$ topological orders give rise to fractonic behaviors in 2D. 
Note that none of the two terms in the perturbation of Eq.~\ref{eq:FTC} can create a pair of anyon, so fractonic behavior is expected. 
At $\alpha = 0$, the TC model has four types of Abelian anyons: $1, e, m, \epsilon = e\times m$, where $e$ and $m$ are electric charge and magnetic flux in the context of $Z_2$ gauge theory, behaving as self bosons and mutual semions; while the $e\times m$ bound state $\epsilon$ have fermionic statistics. The ground state is an anyon vacuum, elementary excitations under open boundary condition can be any one of the $e$, $m$ or $e\times m$ anyons. It is natural to ask what is the dynamics of these elementary particles under an $\alpha \neq 0$ perturbation. Indeed, it is known that the TC, when perturbed by a transverse field $\sum_i \sigma_i^x$ or $\sum_i \sigma_i^z$ \cite{Trebst2007,VidalXZ,mcginley2022}, a single $e$ and single $m$ anyons would disperse isotropically; and by $\sum_i \sigma_i^y$ the fermionic $\epsilon$ particles would disperse linearly \cite{Feng2023,Vidal2009}.  
However, it is clear from the leading order perturbation that in the case of Eq. \ref{eq:FTC}, none of these particles would gain mobility; while it is the anyon bound states, e.g. the two bosonic bound states, $e_1\times e_2$ or $m_1 \times m_2$ with individual anyons aligned in (off)diagonal direction, that develop partially mobile softened modes. The subscript is used to emphasize that they are spacially separated with a lattice constant and do not fuse to a trivial state. It is readily to see that these particles, while mobile, enjoys only a constraint mobility. For example, a bound state created to be aligned with diagonal direction can only move along the off-diagonal direction, as shown in Fig.~\ref{fig:anyon}(left). Nevertheless, the bound state created in vertical or horizontal directions remains stable, as shown in Fig.~\ref{fig:anyon}(right). 
To the leading order perturbation theory, the effective dispersion of an anyon pair, such as that presented in Fig.~\ref{fig:anyon}(left), can be described by
\begin{equation} \label{eq:dispersion}
    \varepsilon_{m_1\times m_2}(\mathbf{k}) = \varepsilon_{e_1\times e_2}(\mathbf{k}) = 4 - 2\alpha \cos(\mathbf{k}\cdot \mathbf{d})
\end{equation}
with $\mathbf{d}$ the primitive lattice vector pointing in the off-diagonal direction. 
Equation~\ref{eq:dispersion} can be readily derived by the Fourier transformation of the perturbed excited state at the first order approximation: at the zeroth order, the $4$ in Eq.~\ref{eq:dispersion} is given by the energy of a composite anyon pair $m_1 \times m_2$ or $e_1 \times e_2$; at the next order, as is shown in Fig.~\ref{fig:anyon}(left), the matrix element of a local $A_p$ perturbation transfers the $m_1 \times m_2$ from the lower left of $A_p$ to the upper right of $A_p$, and vice versa. Let $E_0$ be the ground state energy of $H_{\rm FTC}(\alpha = 0)$, then the first order approximation gives the matrix elements: $H_{\rm FTC}\ket{m_1 \times m_2}_{i} \approx -\ket{m_1 \times m_2}_{i + \mathbf{d}} - \ket{m_1 \times m_2}_{i - \mathbf{d}} + (E_0 + 4)\ket{m_1 \times m_2}_{i}$, whose momentum space representation gives an effective 1D dispersion along the off-diagonal direction described by Eq.~\ref{eq:dispersion}.
Hence the two-anyon bound state does not condense at the transition $\alpha=1$, indeed, the phase transition is instead driven by the four-anyon bound state, which is free to move in all directions. These properties are similar to the 2D fractonic property of the plaquette Ising model (PIM) where four-spin-flip excitation can move without constraint; while single- or two-spin-flip excitations are either immobile or partially mobile along lines \cite{Yan2019a,Vidal2009}.

\subsection{Order parameter}
For the ground state property of Eq.~\ref{eq:FTC} without anyon excitations, we also define the topological order parameter. 
In the large-$\alpha$ limit, a vacuum state of the dual anyons, i.e. $A_p$ and $B_s$ in the other set of TC stabilizers, are guaranteed at ground state by duality. Naturally, we can use these operators, or their product in some (potentially disconnected) region $\mathcal{M}$, as topological order parameters:
\begin{equation} \label{eq:op}
    O_p = \expval{\prod_{p\in \mathcal{M}} A_p},~O_s = \expval{\prod_{s\in \mathcal{M}} B_s}
\end{equation}
such that the phase of $\alpha <(>) 1$ corresponds to $O = 0(1)$. Here we restrict ourselves to open boundary condition, ruling out the non-contractible Wilson loops for convenience. For the smallest $\mathcal{M}$, the order parameter reduces to a single star $A_p$ or plaquette $B_s$.

\section{SSPT model in TC}
Now we present the exact mapping between TPIM, as a strong SSPT model with subsystem symmetries, and the aforementioned FTC model with $Z_2$ topological order. We show explicitly that FTC can be exactly mapped to two copies of the TPIM. The membrane order parameter of the TPIM is exactly mapped to dual TC stabilizers as the order parameter of the frustrated TC model, and SSPT gapless edge states of the TPIM can be mapped exactly to zero-energy dangling operators under open boundaries, and the transition from the SSPT-ordered TPIM to the trivial paramagnetic phase is mapped to the transition between two distinct topological orders. Hence, the energy spectrum of the FTC is effectively a ``double cover'' of the spectrum of the TPIM subjected to subsystem-symmetry-preserving perturbation. 
\subsection{Duality transformation}
\begin{figure*}[t]
\centering
\includegraphics[width=0.95\linewidth]{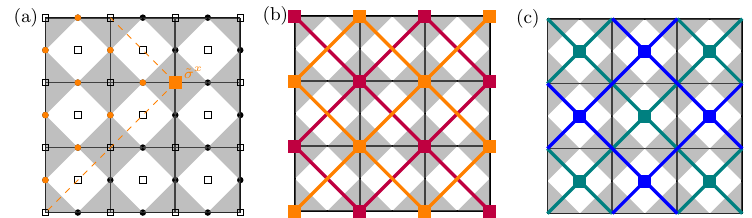}

\caption{(a) An illustration of the transformation in Eq.~\ref{eq:jw}. Qubits in $\Lambda$ lattice are denoted by filled dots, while qubits in the dual lattice $\tilde{\Lambda}$ are denoted by hollow squares. In Eq.~\ref{eq:jw}, the non-local transformation relates a qubit in $\tilde{\Lambda}$ (represented by the orange square) to the set of qubits in $\Lambda$ (denoted by the orange dots) that are located within the dashed triangular region immediately to its left (thus the symbol $>$). (b,c) Two copies of the TPIM in the dual lattices after the transformation of FTC. (b) One copy of the TPIM is defined on vertices in $\tilde{\Lambda}$, denoted by $V$. The two emergent sublattices $V_1, V_2 \in V$ are denoted by filled orange and red squares, corresponding to the two sublattices shown in Fig.~\ref{fig:sum}(right). (c) The second TPIM copy defined on centers of the faces in $\tilde{\Lambda}$, denoted by $F$. The two emergent sublattices $F_1, F_2\in F$ are labeled by filled green and blue squares.  }
\label{fig:trans}
\end{figure*}

Let us now introduce a new set of spin variables living on the vertices and faces of the dual lattice $\tilde{\Lambda}$ \cite{Vidal2009}, where new degrees of freedom are denoted by hollow squares in Fig.~\ref{fig:trans}(a). For the simplicity of the demonstration we assume that the square TC lattice $\Lambda$ is commensurate with the checkerboard lattice such that the map to $\tilde{\Lambda}$ is consistent in both bulk and boundaries. We apply a non-local transformation, relating a pauli matrix $\tilde{\sigma}_x$ on the dual lattice to a set of qubits on the original lattice, and identify $\tilde{\sigma}^z$ as stabilizer operators:
\begin{equation} \label{eq:jw}
	\tilde{\sigma}_j^x = \prod_{i>j} \sigma_i^y,~~ \tilde{\sigma}_{j(s)}^z = A_s,~~ \tilde{\sigma}_{j(p)}^z = B_p
\end{equation}
as is illustrated in Fig.~\ref{fig:trans}(a), $\tilde{\sigma}_j^x$ on the dual lattice (orange square) is given by the product of $\sigma_i^y$ on the original lattice (orange dots). It is readily to check that it is a faithful representation that respects Pauli algebra. Under this mapping, one copy of the TC can be treated as trivial Ising variables, while the other is responsible for all emerging non-trivial properties. Note that this map preservers the number of degrees of freedoms, thus no additional constraint or projection is needed. 
Using $(\sigma^y)^2 = 1$, it is readily to see that the non-local mapping in Eq.~\ref{eq:jw} [Fig.~\ref{fig:trans}(a)] essentially defines the following duality map:
\vspace{-1em}
\begin{equation} \label{eq:pic}
    \TikzFigOne \longleftrightarrow \;\;\;\TikzFigTwo
\end{equation}
that is, the product of the four $\tilde{\sigma}^x$'s on the left hand side is mapped to a single $\sigma^y$ on the right hand side, and vice versa.
Using $(\tilde{\sigma}^x)^2 = 1$, it is readily to see that the product of four adjacent $\sigma^y$'s is mapped into the product of four next-nearest $\tilde{\sigma}^x$'s. 
\begin{widetext}
    \begin{equation} \label{eq:pic2}
    \TikzFigYcheck \;\; \leftrightarrow \;\; \TikzFigThree{0} \;\; \times \;\;\; \TikzFigThree{90} \;\; \times\;\;
    \TikzFigThree{180} \;\;\ \times \;\; \TikzFigThree{270} \;\; = \;\; \TikzFigXXXX
\end{equation}
\end{widetext}
where we kept the same convention used in Eq.~\ref{eq:pic}, i.e. connections between gray bullets denote exchange interaction of $\sigma^y$; and connections between squares denote that of $\tilde{\sigma}^x$.  
Hence, Eq.~\ref{eq:pic2} and Eq.~\ref{eq:jw} relate the checkerboard lattice and its dual lattice by the following dual relation:
\begin{equation} \label{eq:duals}
    \prod_{i\in \rsquare}\tilde{\sigma}_i^x \leftrightarrow {\sigma}^y_{c(\rsquare)},~~ \tilde{\sigma}_{i\in \tilde{\Lambda}}^z \leftrightarrow A_s \cong B_p
\end{equation}
where $i\in \rsquare$ means the four sites of one of the smallest rhombuses in $\tilde{\Lambda}$ consisting of two vertex squares and two face-center squares in shown in Fig.~\ref{fig:trans}(a), and $c(\rsquare)$ denotes the site in $\Lambda$ that lies at the center of the rhombus $\rsquare$ in $\tilde{\Lambda}$ (see also Eq.~\ref{eq:pic}).
As we are about to show, this map reveals an interesting connection between the dual TC stabilizers (Eq.~\ref{eq:dualtc}) and TPIM stabilizers. For clarity, let us first rewrite the operators of the dual TC, $B_s$ and $A_p$, in terms of the original TC variables
\begin{equation} \label{eq:bsap}
    B_s = A_s \prod_{i\in s} \sigma_i^y,~~
    A_p = \prod_{i\in p}\sigma_i^y \:B_p
\end{equation}
simply by the identity $\sigma^a \sigma^b = \delta_{ab} + i\epsilon^{abc} \sigma^c$. 
Use the duality of $\Tilde{\sigma}^z \leftrightarrow A_s (B_p)$ in Eq. \ref{eq:jw} or Eq.~\ref{eq:duals}, and apply Eq.~\ref{eq:pic2} for $\prod_{i\in s(p)}\sigma_i^y$, we see that $A_p$ and $B_s$ of Eq.~\ref{eq:bsap} can be mapped into: 
\begin{equation} \label{eq:jwd}
    B_s \leftrightarrow \tilde{\sigma}^z_{c(\tilde{p}_V)} \prod_{i\in \tilde{p}_V} \tilde{\sigma}_i^x,~~ 
    A_p \leftrightarrow\tilde{\sigma}^z_{c(\tilde{p}_F)} \prod_{i\in \tilde{p}_F} \tilde{\sigma}^x_i
\end{equation}
Here $\tilde{p}_V$ denotes a plaquette defined on the vertices of the dual lattice $\tilde{\Lambda}$ that encloses one vertex in $\tilde{\Lambda}$, as illustrated in Fig.~\ref{fig:trans}(b), where the enclosed qubit and the qubits of the plaquette live in different emergent sublattices, colored by orange and purple squares.  $\tilde{p}_F$ denotes a plaquette defined on face centers of $\tilde{\Lambda}$ that encloses another face center in $\tilde{\Lambda}$, as illustrated in Fig.~\ref{fig:trans}(c). Again the enclosed qubit and the qubits of the plaquette live in different emergent sublattices, colored by green and blue squares. 
$c(\tilde{p}_V)$ denotes the site that lies at the center of a vertex-plaquette, while $c(\tilde{p}_F)$ denotes that of a face-center plaquette. 
After this mapping, it is readily to see that the FTC amounts to adding up interactions in $V,F \in \tilde{\Lambda}$: 
\begin{equation} \label{eq:csh}
\begin{split}
        \mathcal{H}_\eta = -\alpha \sum_{c(\tilde{p}_\eta)} \tilde{\sigma}^z_{c(\tilde{p}_\eta)} \prod_{i\in \tilde{p}_\eta} \tilde{\sigma}_i^x 
    - \sum_{c(\tilde{p}_\eta)} \tilde{\sigma}^z_{c(\tilde{p}_\eta)},~~ \eta = V,F
\end{split}
\end{equation}
where each $\eta$, represented in Fig.~\ref{fig:trans}(b) or (c), corresponds to one copy of the TPIM. 
We use the calligraphic $\mathcal{H}$ to remind the reader that it is one of the two disjoint TPIM Hamiltonians. 
Eq. \ref{eq:csh} is equivalent to the TPIM (or cluster-state model) protected by a subextensive number of $Z_2 \times Z_2$ linear symmetries. For $\alpha < 1$, Eq.~\ref{eq:csh} reduces to Ising paramagnet in $\tilde{\Lambda}$ which is equivalent to a TC; while when $\alpha > 1$ the system becomes strong SSPT state. 
It is also clear that since the two competing topological orders in Eq.~\ref{eq:FTC} cannot be connected to each other via local operations, the TPIM model at the large $\alpha$ limit of Eq.~\ref{eq:csh} cannot be continuously deformed into the trivial product state at small $\alpha$.
For each copy, e.g. one defined on the $V$ lattice, the symmetry generators are given by:
\begin{equation} \label{eq:symm}
    U_m^{V_i} = \prod_{n \in\rm cols} \tilde{\sigma}_{m,n}^z,~ U_n^{V_i} = \prod_{m \in\rm rows} \tilde{\sigma}_{m,n}^z
\end{equation}
The notation $U_{m(n)}^{V_i}$ stands for the linear unitary operators defined on each row (column) inside the emergent sublattice $V_i$, where $i = 1,2$ corresponds to the purple and orange sites in Fig.~\ref{fig:trans}(b), 
and the $m,n$ labels the rows and columns of $V_i$. 
It is readily to see that each of the symmetry generators corresponds to the product of canonical TC stabilizers along a (off)diagonal line, which is also a symmetry of the dual TC Hamiltonian.  
Assume the original lattice $\Lambda$ is of the size $2L_x \times 2L_y$ ($L_x \times L_y$ unit cells in checkerboard lattice), Eq. \ref{eq:csh} describes two copies of the TPIM model, with each defined on a $L_x \times L_y$ sublattice having
\begin{equation} \label{eq:gsd}
    {\rm GSD}_{V(F)} = 4^{L_x + L_y - 1}
\end{equation}
corresponding to the subextensive $Z_2 \times Z_2$ symmetry-protected edge modes of the TPIM. 
Indeed, one may choose to perturb the TC Hamiltonian only by either one of $A_p$ or $B_s$ to get only one copy of the TPIM, and a lattice of $L_x\times L_y$ would give the same GSD. 
The GSD due to protected edge states can be related to the dangling edge operators in the TC lattice under OBC, as we will discuss in the next section.

In fact, a closer inspection of Eq.~\ref{eq:jwd} and Eq.~\ref{eq:csh} readily reveals that a single TPIM can emerge from a half of the FTC, whereby only one of $V$ or $F$ sublattices emerge:
\begin{equation} \label{eq:ssh}
\begin{split}
         -\sum_s \left(A_s + \alpha B_s\right) \cong - \sum_p\left(B_p +\alpha A_p\right) \mapsto H_{\rm TPIM}
\end{split}
\end{equation}
In this form the perturbation analysis of anyon mobility is no longer viable, 
while it singles out a minimal stabilizer model that is mappable to a strong SSPT-ordered TPIM in Eq.~\ref{eq:csh}.

\subsection{TC stabilizer as dual SSPT order parameter}
Given the correspondence between SSPT states and the FTC model, it would be interesting to relate the order parameters in the two cases.  
While there is no local order parameter (local in $\tilde{\Lambda}$ basis) for distinguishing the small-$\alpha$ and large-$\alpha$ phases of Eq. \ref{eq:csh}, it is known to exist a two-dimensional membrane order parameter $\tilde{O}$ \cite{You2018}:
\begin{equation} \label{eq:op2}
    \tilde{O} = \expval{\prod_{i\in \mathcal{C}} \tilde{\sigma}_i^x \prod_{i\in \mathcal{M}} \tilde{\sigma}_i^z }
\end{equation}
where $\mathcal{C}$ refers to the four corners of a square membrane of one sublattice, and $\mathcal{M}$ refers to all sites of the other sublattice which are enclosed by $\mathcal{C}$. It immediately follows that $\tilde{O} \mapsto O_{p,s}$, i.e.
$\tilde{O}$ can be mapped, according to Eq. \ref{eq:jw}, to the product of disconnected stabilizer operators as defined in Eq. \ref{eq:op}, which serves as a natural order parameter of the FTC in Eq. \ref{eq:FTC}. In particular, if $\mathcal{C}$ in Eq. \ref{eq:op2} is chosen to be the smallest possible membrane, i.e. that which contains only one site of the other sublattice, $\tilde{O}$ would be mapped to a single $A_p$ or $B_s$ of the dual TC. Such correspondence between FTC and SSPT provides a lucid intuition for the topological nature of strong SSPT states, that the membrane order in SSPT cluster states is related to Abelian anyon excitations of a TC; or conversely, one may start from a TC model, where dressing an anyon operator with non-local string operators will give rise to a membrane topological order in SSPT cluster states.

\subsection{Dangling edge operators and SSPT edge modes} \label{sec:edge}
To test the validity of the aforementioned duality mapping between TPIM and the FTC model, we study the fate of SSPT edge symmetries and its resultant degeneracy under the mapping. 
Just like SPT phases of matter where global symmetries are projectively realized on edges, in the ground state sector of the TPIM, the subsystem symmetry groups are projectively realized on the two edges of each subsystem, where gapless spin-$\frac{1}{2}$ degrees of freedoms are protected by subsystem symmetries. An alternative way to understand the GSD of edge modes is to map the TPIM to a SSB model, where the GSD of symmetry-broken ground states corresponds to edge mode degeneracy of the TPIM \cite{You2018}. While it provides a clear way to understand the GSD and the fact that SSPT states of the TPIM cannot be continuously deformed into trivial state while preserving symmetries, such mapping to an SSB model completely hides the edge physics in the bulk. 
In the dual picture, as we shall see, the edge modes are manifested in terms of dangling edge operators that do not cost energy, providing an alternative yet lucid picture of the subextensive degeneracy of SSPT edge states.   

\begin{figure}[t]
\centering
\includegraphics[width=\linewidth]{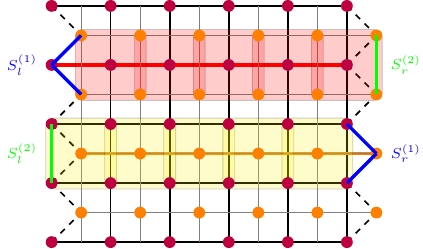}

\caption{An illstration of the projectively realized subsystem symmetries $U_m = \prod_{i} \tilde{\sigma}^z_{m,i}$ (product of $\tilde{\sigma}^z$'s on the horizontal red line) and $U_n = \prod_{i} \tilde{\sigma}^z_{n,i}$ (product of $\tilde{\sigma}^z$'s on the horizontal orange line). on a finite lattice with OBC in the horizontal direction. Dashed lines mark out the zigzag boundaries consisting of spins of alternating sublattice. TPIM stabilizers marked in transparent plaquettes projectively realize the subsystem symmetries $U_m^V$ or $U_n^F$ as edge symmetry operators $S^{(1)}$ and $S^{(2)}$ in the ground state sector. }
\label{fig:edgeop}
\end{figure}

\begin{figure*}[t]
\centering
\includegraphics[width=0.9\linewidth]{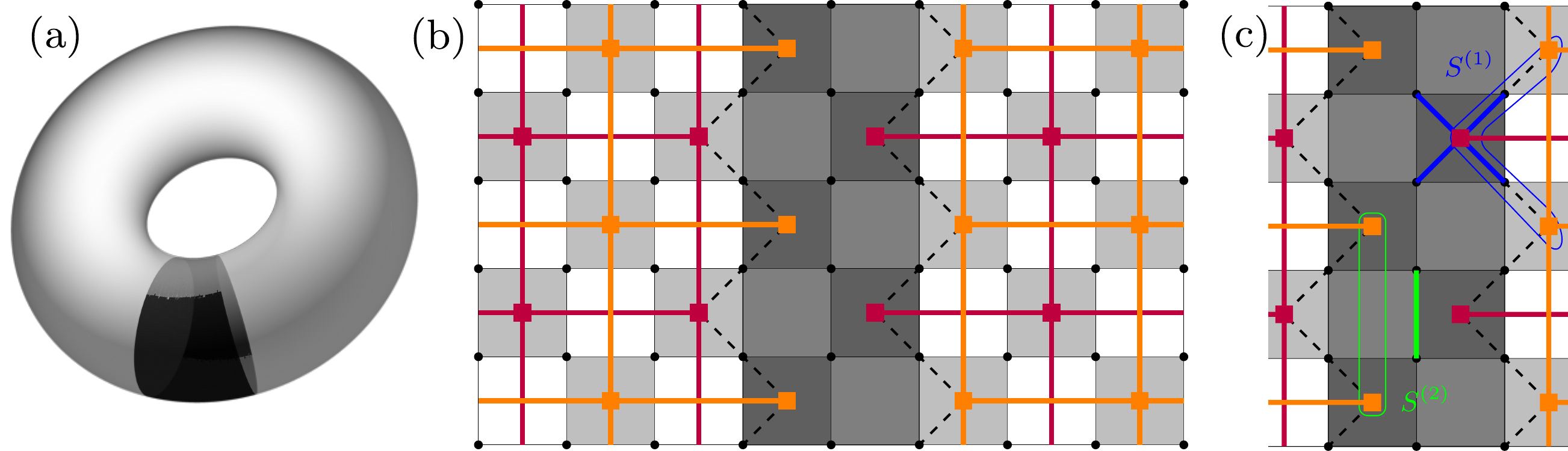}
\caption{Open boundaries and edge symmetries of dual TC. (a) Illustration of an open boundary FTC model where two boundaries are realized by removing columns of stabilizers (black strip) from the Hamiltonian. (b) The rotated TC lattice parallel to Wen's plaquette lattice. For clarity we only present $V \in \tilde{\Lambda}$, corresponding to the TPIM defined on the vertices of the dual lattice shown in Fig.~\ref{fig:trans}(b). The removed stabilizer interactions are represented by the two shaded black columns corresponding to the black strip in (a). Dashed zigzag lines denote the two boundaries of the TPIM model.  (c) The symmetry operators on the edges of the TPIM. Under the duality map defined in Eq.~\ref{eq:jw}, the edge symmetry $S^{(1)}$ (thin blue line) is mapped to the stabilizer of the dual TC $B_s$ (thick blue line), and $S^{(2)}$ symmetry (thin blue line) to $\sigma_i^y \sigma_{i+\hat{y}}^y$ on the boundary (thick green line).  }
\label{fig:edge}
\end{figure*}
Before we delve into the mapping between edge operators in $\Lambda$ and SSPT edge modes in $\tilde{\Lambda}$, we first briefly discuss the fractionalized subsystem symmetries in TPIM and its resultant edge degrees of freedoms. 
Let us first scrutinize the ground state $\ket{\psi_0}$ of one copy of the TPIM corresponding to a rotated Fig.~\ref{fig:trans}(b), as is presented in Fig.~\ref{fig:edgeop}. For convenience we denote spins in $V_1$ sublattice as $\tilde{\sigma}$, spins in $V_2$ sublattice as $\tilde{\tau}$, and the two types of the TPIM stabilizers as: $A_c \equiv \tilde{\tau}_{c}^z\prod_{i\in \rm corners} \tilde{\sigma}_{i}^x$ (light red plaquettes in Fig.~\ref{fig:edgeop}), $B_c \equiv \tilde{\sigma}_{c}^z\prod_{i\in \rm corners} \tilde{\tau}_{i}^x$ (light yellow plaquettes in Fig.~\ref{fig:edgeop}), where $\tilde{\sigma}$ and $\tilde{\tau}$ belong to different sublattices. Similar to 1D SPT models \cite{verresen2017one}, these subsystem symmetries can be understood as gauge operators in the ground state for they commute with the Hamiltonian. By such gauge symmetry we have
\begin{equation}
    \ket{\psi_0} = A_1 A_2 \cdots A_{L} \ket{\psi_0} = B_1 B_2 \cdots B_{L} \ket{\psi_0}
\end{equation}
as is illustrated by the joint sequence of plaquettes in Fig.~\ref{fig:edgeop}, 
so that the global symmetry operators are equivalent to
\begin{align}
    U_m^{V_1} \ket{\psi_0} &= \prod_{i\in A} \sigma_i^x \prod_{j\in \rm V_{1,\rm row}}^{L}A_j \ket{\psi_0} = S^{(1)}_l S^{(2)}_r \ket{\psi_0} \label{eq:uap}\\
    U_n^{V_2} \ket{\psi_0} &= \prod_{i\in B} \tau_i^x \prod_{j\in \rm V_{2, \rm row}}^{L} B_j \ket{\psi_0} = S^{(2)}_l S^{(1)}_r \ket{\psi_0} \label{eq:ubp}
\end{align}
where $U_m^{V_1}$ and $U_n^{V_2}$ are two subsystem symmetries of $V_1$ and $V_2$ sublattices as defined in Eq.~\ref{eq:symm}, and the edge operators $S^{(1),(2)}_{l,r}$ are given by

\begin{align}
    S^{(1)}_l &= \tilde{\tau}^x_i \tilde{\sigma}^z_{i+1} \tilde{\tau}^x_{i+2},~ 
    S^{(2)}_l = \tilde{\sigma}^x_i \tilde{\sigma}_{i+1}^x \label{eq:s1}
    \\ 
    S^{(1)}_r &= \tilde{\sigma}^x_i \tilde{\tau}^z_{i+1} \tilde{\sigma}^x_{i+2},~
    S^{(2)}_r = \tilde{\tau}^x_i \tilde{\tau}_{i+1}^x \label{eq:s2}
\end{align}
Thus the action of the subsystem symmetries can be factored into operations acting on the left and right edges separately. Such picture of symmetry fractionalization also applies to vertical open boundaries by the same token, as well as other edge geometries \cite{You2018}. 
Too see explicitly the edge degrees of freedoms and the resultant ground state degeneracy, let us define the following operators on one of the edges. Here we use the left edge as an example.
\begin{equation}
    \pi_i^z = \tilde{\tau}^x_{i} \tilde{\sigma}^z_{i+1} \tilde{\tau}^x_{i+2},~ \pi_i^y = \tilde{\tau}^x_i \tilde{\sigma}^y_{i+1} \tilde{\tau}^x_{i+2},~\pi_i^x = \tilde{\sigma}^x_i
\end{equation}
these operators commute with the bulk stabilizers of the TPIM, and satisfy the Pauli algebra on the left edge of Fig.~\ref{fig:edgeop}. The projective symmetries on the left edge can thus be written as $S^{(1)}_{i,l} = \pi_i^z$ and $S^{(2)}_{i,l} = \pi_i^x \pi_{i+1}^x$. For a boundary with linear size $L$, these form $L$ spin-$\frac{1}{2}$ degrees of freedoms on the edge, giving rise to $2^L$-fold degeneracy. Considering a lattice with OBC in both directions, we arrive at the GSD given in Eq.~\ref{eq:gsd}. Since there are no additional local symmetric terms that can be added to the TPIM, the edge modes are protected by subsystem symmetries.

Now it is readily to see that the edge degrees of freedoms of the TPIM is neatly reflected in the zero-energy edge operators in the FTC model revealed by the duality. Consider FTC model with cylindrical geometry where its two boundaries are realized by removing two columns of (dual) TC stabilizers on a torus, as illustrated in Fig.~\ref{fig:edge}(a,b). Note that while the TC stabilizer interactions in the black-shaded region are removed, the spins thereof are kept so that they can be mapped to edge spins in $\tilde{\Lambda}$ of the TPIM. This is the most compact construction of OBC without inducing redundant degrees of freedoms -- removing more than two columns of stabilizers would lead to redundant spins in $\Lambda$ that do not correspond to any spins in $\tilde{\Lambda}$; whereas removing a single column of stabilizers would cause conflict in assigning spins of $\Lambda$ to boundary spins of $\tilde{\Lambda}$. The corresponding degrees of freedoms in $\tilde{\Lambda}$ is denoted by filled purple or orange squares in Fig.~\ref{fig:edge}(b) where we recover the same edge geometry for TPIM as Fig.~\ref{fig:edgeop} after aforementioned duality mapping. 
By Eq.~\ref{eq:jw}, the edge symmetry operators $S^{(1)}$ and $S^{(2)}$ in TPIM, as given in Eq.~\ref{eq:s1} and Eq.~\ref{eq:s2}, are mapped into local operators
\begin{equation}
    S^{(1)} \mapsto B_s,~
    S^{(2)} \mapsto \sigma^y_{i} \sigma_{i+1}^y
\end{equation}
which are supported only on the black-shaded region in Fig.~\ref{fig:edge}(b,c). 
An illustration of these symmetry operators in presented in Fig.~\ref{fig:edge}(c), details for the above mapping is shown in Appendix \ref{sec:app}. Since $B_s$ and $\sigma^y_{i} \sigma_{i+1}^y$ are located in the shaded region of Fig.~\ref{fig:edge}(b,c) where all stabilizer interactions are removed, the operation contribute no energy, resulting in the $2^L$-fold degeneracy on the edge of one copy of the two TPIMs. 


Given such correspondence between edge symmetries of the TPIM and dangling edge operators of FTC, it is easy to understand the degeneracy due to SSPT boundary modes in terms of stabilizers, and thereby demonstrate the validity of the aforementioned duality mapping between TPIM and the FTC model. 
Suppose the checkerboard lattice of the FTC model is of a linear size $2L$ ($L$ unit cells). Setting the stabilizer interactions within the shaded region of Fig.~\ref{fig:edge}(a,b) to zero would leave $4L$ stabilizers unconstrained, giving a totality of $2^{4L}$ GSD with each of the two copies of the TPIMs getting ${\rm GSD} = 2^{2L}$. For the TPIM copy in  $\tilde{\Lambda}_{V}$, for instance, it corresponds to $L$ left boundary modes and another $L$ right boundary modes. We can further set up a FTC model with OBC in both directions by removing another two rows of TC stabilizers along the big circle of the torus, in analogy to Fig.~\ref{fig:edge}(a). This results in $8L - 4$ unconstrained stabilizers (assume $L_x = L_y = L$), where the $-4$ is due to the doubly counted stabilizers on the cross between two non-contractible loops on the torus. Hence, each of the two copies of the TPIM gets its share of ${\rm GSD} = 2^{4L - 2}$, which is consistent with Eq.~\ref{eq:gsd} assuming $L_x = L_y = L$.



\subsection{Relation to the plaquette Ising model}
To fully elucidate the physical understanding, it is imperative to investigate the relationship between the plaquette Ising model (PIM) \cite{Xu2004,Xu2005} and TC. Notably, a TPIM has been known to map onto two copies of PIM, leading to the designation of the former as the topological PIM (TPIM) \cite{You2018}. To see the relation between TC and PIM and TPIM, 
consider the following Hamiltonian as an extended TC with additional ``Y-checks'', which we shall refer to as YTC:
\begin{equation} \label{eq:tcy}
    H_{\rm YTC} = -\sum_s A_s - \sum_p B_p - \alpha \sum_s Y_s - \alpha \sum_p Y_p
\end{equation}
Here we defined $Y_{s(p)} = \prod_{i\in s(p)} \sigma_i^y$, which does not commute with the $A_s$ nor $B_p$. 
At small $\alpha$, single anyon $e$, $m$ and their fermionic bound states $\epsilon = e\times m$ are either completely immobile or partially mobile along lines, and only bosonic bound states $e_1 \times e_2$ or $m_1 \times m_2$ are completely mobile whose dispersion is again described by Eq. \ref{eq:dispersion}. This perturbation picture of anyon mobility is similar to the case of FTC, however, as we are to discuss, the YTC is mappped to PIM, instead of SSPT-ordered TPIM. 
The perturbation is Eq.~\ref{eq:tcy} is equivalent to a rotated Xu-Moore model or PIM, however, 
following the same map in Eq. \ref{eq:jw}, it is straightforward to see Eq. \ref{eq:tcy} is mapped to \emph{four copies} of PIM under a Zeeman field
\begin{equation} \label{eq:pim}
    \mathcal{H}_\xi = -\alpha \sum_{\tilde{p}_\xi} \prod_{i\in \tilde{p}_\xi} \tilde{\sigma}^x_{i} - \sum_{i\in \tilde{\Lambda}_\xi} \tilde{\sigma}_i^z
\end{equation}
where the subscript $\xi$ in $\tilde{p}_\xi$ and $\tilde{\Lambda}_\xi$ corresponds to the four distinct sublattices $\xi = 1,\cdots,4$ of  $\tilde{\Lambda}$. This is the PIM that also enjoys a duality $\alpha \leftrightarrow \alpha^{-1}$, thus a transition at $\alpha = 1$. Indeed, as is discussed in \cite{Vidal2009,Feng2023}, a Zeeman field in $\sigma^y$ would induce a single copy of PIM after the transformation, where there is no sublattice structure in $\tilde{\Lambda}$. 
Each sublattice Hamiltonian in Eq.~\ref{eq:pim} enjoys a subextensive number of $Z_2$ symmetries generated by $U_m^\xi = \prod_{n \in\rm cols} \tilde{\sigma}_{m,n}^z,~ U_n^\xi = \prod_{m \in\rm rows} \tilde{\sigma}_{m,n}^z$, 
with $m,n$ are row and column indices bounded by the lattice size. A direct count yields a GSD for each pair of replicas equivalent to that of a TPIM, assuming the same lattice size characterized by either $V$ or $F$. 

We have explicated the relation between the YTC and PIM, but what is the relation between YTC and FTC regarding the hidden SSPT phase in the later? 
Consider two of the four sublattices $\xi = 1,2$, and define a non-local transformation: $\bar{\sigma}_{i_1}^{x} = \tilde{\sigma}_{i_1}^x \prod_{i_2\ge i_1} \tilde{\sigma}_{i_2}^z$, $\bar{\sigma}_{i_2}^{x} = \tilde{\sigma}_{i_2}^x \prod_{i_1\ge i_2} \tilde{\sigma}_{i_1}^z$, $\bar{\sigma}_{i_{1(2)}}^z = \tilde{\sigma}_{i_{1(2)}}^z$, 
where $i_1\in \tilde{\Lambda}_{\xi_1}$ and $i_2\in \tilde{\Lambda}_{\xi_2}$. 
It is straightforward to check that under these transformation, the combination of the two copies in Eq.~\ref{eq:pim} becomes the TPIM Hamiltonian in Eq.~\ref{eq:csh} in $\bar{\sigma}$ variables.  
Here one may immediately recognize that the YTC, as the mapped model of PIMs, can be converted to FTC, as the mapped TPIMs, by simply hitting all $Y_s$ and $Y_p$ operators in Eq.~\ref{eq:tcy} by $A_s$ and $B_p$.
Hence the non-local transformation between TPIM and PIM in the dual lattices is achieved locally by applying canonical TC stabilizers, further establishing the deep relation between SSPT phases and the TC topological order. A summary of these mappings is presented in Table.~\ref{table}.

\begin{figure}[t]
\centering
\includegraphics[width=\linewidth]{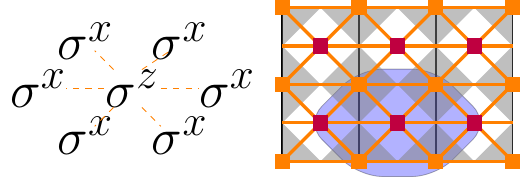}
\caption{The triangular cluster model with SSPT order. The generator of the SSPT model is the seven-point interaction shown on the left side and the blue shadow on the right side. It can be mapped onto a frustrated 2D TC model defined in Eq.~\ref{eq:FTC2}, whereby a triangular lattice in the dual lattice merge. Here for clarity we label the face-centered sites of $\tilde{\Lambda}$ in red, and vertex sites of $\tilde{\Lambda}$ in orange. }
\label{fig:nat}
\end{figure}
\section{Map to triangular cluster model}
We have established a mapping from FTC to two copies of the TPIM, as illustrated in Fig.~\ref{fig:trans}(b) and (c). A natural question to ask is if this insight would be useful in the understanding of other SSPT models. Indeed, a lot of different 2D SSPT models have been proposed using symmetry defect homology theory \cite{Nathanan2020}. The answer is affirmative. Here we give an example of mapping another frustrated TC model into the SSPT cluster model on a triangular lattice, whose Hamiltonian is generated by seven-qubit interactions as shown in Fig.~\ref{fig:nat}(left). This model has strong SSPT order as is proved in Ref.~\cite{Yizhi20180}.  Consider the following frustrated TC Hamiltonian that is different from Eq.~\ref{eq:FTC}:
\begin{equation} \label{eq:FTC2}
\begin{split}
        &H_{\rm FTC_2} = -\sum_s A_s - \sum_p B_p  \\ 
        &+ \alpha \sum_{p} \sigma_{1,p}^z \sigma_{2,p}^x \sigma_{3,p}^z \sigma_{4,p}^x + \alpha \sum_{s} \sigma_{1,s}^x \sigma_{2,s}^z \sigma_{3,s}^x \sigma_{4,s}^z
\end{split}
\end{equation}
where the symbol $\sigma_{1,s(p)}^{z(x)}$ denotes the $i$-th spin in a plaquette or star. The index is increased clock-wisely and the bottom site is assumed to be $i=0$. 
The perturbation term, up to a unitary rotation, is equivalent to an antiferromagnetic Xu-Moore model or the plaquette Ising model \cite{Xu2004}. 
It does not commute with $H_{\rm TC}$ hence the topological order is again frustrated. For clarity we rewrite the perturbation term into 
\begin{align}
     &\sigma_{1,p}^z \sigma_{2,p}^x \sigma_{3,p}^z \sigma_{4,p}^x = -\sigma_{2,p}^y \sigma_{4,p}^y B_p \\
     &\sigma_{1,s}^x \sigma_{2,s}^z \sigma_{3,s}^x \sigma_{4,s}^z = -\sigma_{2,s}^y \sigma_{4,s}^y A_s
\end{align}
Applying the transformation in Eq.~\ref{eq:jw} to the right hand side gives us the cluster model defined on a triangular lattice, as is illustrated in Fig.~\ref{fig:nat}. 
Indeed, this is similar to the Xu-Moore duality discussed in \cite{Nathanan2020,Xu2004} where the triangular cluster model is dual to Wen's plaquette model, however, since here we used a different map, the triangular cluster model is mapped to a rotated Xu-Moore model. The fact that the (rotated) Xu-Moore model has a subsystem symmetry-breaking order also makes it clear that the triangular cluster model in Fig.~\ref{fig:nat} is SSPT ordered -- there is no local finite-depth local unitary operation that can connect the symmetry-breaking Xu-Moore model to the topologically ordered TC, thus the triangular cluster model, as the dual model of the former, cannot be continuously deformed into a trivial product state as the dual model of the later. Again the order parameter of Eq.~\ref{eq:FTC2} can be chosen as the TC stabilizer, which can be $\pm 1$ for large or small $\alpha$, distinguishing the SSPT order from the trivial state in the dual lattice.  
Other SSPT models with higher-order interactions can be generated by the same token, which we do not enumerate in the paper.

\section{Discussion and conclusion}
Given the aforementioned mappings, we briefly comment on the differences between weak SSPT phases, which can be viewed as stacked or weakly coupled 1D SPT phases that retain their individual symmetries, and strong SSPT phases, which cannot be reduced to merely decoupled or weakly coupled 1D SPTs. It's useful to consider the hidden symmetry-breaking picture in SPT phases. Take for instance a standard 1D SPT with $Z_2 \times Z_2$ symmetry and its corresponding non-local string order parameter. It is related to the symmetry-breaking picture of a hidden antiferromagnetic model after a non-local transformation \cite{Kennedy1992a,Kennedy1992b,Pollmann2012}. On the other hand, under the new mapping we introduced in Eq.~\ref{eq:jw}, the TPIM can be mapped to two competing topological TC models in Eq. \ref{eq:FTC}; or equivalently, a Half FTC-stabilizer model, corresponding to a set of mutually frustrating TC stablizers; and the membrane order parameter associated with TPIM becomes a product of stabilizer operators in TC, underlining its intrinsic many-body nature as opposed to weak SSPT systems. This also makes explicit the fact that the SSPT-ordered TPIM model cannot be continuously deformed into a trivial phase in the thermodynamic limit, and the self-duality of the TPIM is made manifestly clear in terms of the self-dual frustrated TC model. 
Interestingly, such connection by duality not only exists between TPIM and TC model, but also holds for other SSPT models such as the triangular cluster model, which can emerge by inducing a different perturbation on TC, supporting the construction of SSPT order using defect homology theory and Xu-Moore duality from a different angle \cite{Nathanan2020}.

\begin{table}[t]
\centering
\setlength{\tabcolsep}{2mm} 
\caption{Variants of TC defined on $\Lambda$ lattice (before transformation) and the effective model of its map in the $\tilde{\Lambda}$ lattice (after transformation)}
\begin{tabular}{|c|c|}
\hline
$\sigma \in \Lambda$  & $\tilde{\sigma}\in \tilde{\Lambda}$  \\ \hline
TC & Paramagnet \\ \hline
TC with $\sigma^y$ field  & PIM \\ \hline
YTC (Eq.~\ref{eq:tcy}) & PIM $\times 4$ \\ \hline
FTC (Eq.~\ref{eq:FTC}) & TPIM $\times 2$ (SSPT) \\ \hline
Half FTC (Eq.~\ref{eq:ssh}) & TPIM (SSPT) \\ \hline
FTC$_2$ (Eq.~\ref{eq:FTC2}) & Triangular cluster model (SSPT) \\ \hline
\end{tabular} \label{table}
\end{table}

In summary, we have constructed exact maps connecting frustrated TC models to different SSPT-ordered models, showing that the novel SSPT phases of matter can be understood in terms of the simplest topological order in TC model.
Given the rapidly evolving theories and experiments in quantum stabilizers and measurement-induced dynamics in various platforms \cite{sagar2022,sriram2022,Giulia2022,Nat2023}, 
it is both theoretically and practically appealing that one may use the frustrated TC-stabilizer measurement to simulate SSPT phases and its fractonic dynamics, and construct equivalences of different SSPT models as dressed TC models. A recent study has shown the possibility of stabilizing measurement-induced SPT phases of 1D cluster Hamiltonians \cite{Yepes2023}. It is intriguing to generalize this framework into 2D SSPT phases of matter. In particular, given the intimate relation between TC and TPIM presented in this work, it is promising that one may use the well-established TC model and relevant platforms to conduct measurement-induced phase transitions between SSPT and trivial phases. Moreover, the exact map between TPIM and FTC suggests that the energy spectrum of the FTC is effectively a ``double cover'' of the spectrum of the TPIM subjected to subsystem-symmetry-preserving perturbation, hence, it establishes the possibility of simulating the thermodynamics of the TPIM using recently realized TC model in cold atom setups \cite{Semeghini2021}.  
Since the TPIM model was originally proposed as a pathway to the one-way quantum computer \cite{Briegel2001b,Briegel2003}, it is intriguing to relate the measurement-based quantum computation to quantum error correction formalism.  
We believe that the our results which relate SSPT phase and the TC stabilizers is suggestive enough to warrant further studies.

{\bf {Acknowledgement:}}
S. Feng acknowledges support from the Presidential Fellowship at The Ohio State University. S. Feng also thanks the Boulder Summer School 2023 at the University of Colorado Boulder for the enlightening lectures and discussions. Further gratitude is extended to X. Yang, N. Trivedi, A. Agarwala, and S. Bhattacharjee for their discussions and collaboration on related topics.

\appendix
\section{Edge operators} \label{sec:app}
In Sec.~\ref{sec:edge} we have shown the correspondence between edge symmetries of the TPIM and dangling edge operators of FTC under OBC, making the SSPT boundary modes explicitly clear in terms of zero-energy operators in TC's degrees of freedoms. Here we present details of its derivation.  Again we consider a cylindrical geometry shown in Fig.~\ref{fig:edge} where two columns of TC interactions are removed. This effectively creates an open boundary in $\Lambda$, where the middle vertical line of the shaded region [Fig.~\ref{fig:edge}(b)] can be viewed as the right boundary of the left-half torus and the left boundary of the right-half torus. 
Note that on the torus with such a stabilizer cut, it is necessary to modify the non-local mapping in Eq.~\ref{eq:jw}, such that the region of the non-local product for the dual representation of $\tilde{\sigma}^x$ [Fig.~\ref{fig:trans}(a)] extends towards its nearest boundary, and the product of $\sigma^y$'s terminates on the boundary. It is readily to see that such modification conserves the Pauli algebra and the duality relation defined in Eq.~\ref{eq:duals}.

We now show that, when mapped into $\Lambda$, the degrees of freedoms on the zigzag boundary of the TPIM [that of $\tilde{\Lambda}$, Fig.~\ref{fig:edge}(b,c)] give rise to the spins that lie in the black-shaded region of $\Lambda$. First we consider $S^{(2)}_r = \tilde{\tau}^x_i \tilde{\tau}_{i+1}^x$ defined in Sec.~\ref{sec:edge}. Below we use a pictorial representation, where solid green squares denote $\tilde{\tau}^x \in \tilde{\Lambda}$, and solid green bullets denote $\sigma^y \in \Lambda$. Hence, using the fact that $(\sigma^\alpha)^2 = 1,\;\forall \alpha = x,y,z$, the edge operators $S^{(2)}_r$ can be mapped according to the following diagrammatic approach:
\begin{widetext}
    \begin{equation}
        S^{(2)} = \tilde{\tau}_i^x \tilde{\tau}_{i+1}^x \;\equiv\;\; \TikzAppa \;\; \leftrightarrow \;\; \TikzAppb \;\;\times \;\;\TikzAppc \;\; = \;\; \TikzAppd = \sigma_{i}^y \sigma_{i+1}^y
    \end{equation}
\end{widetext}
where the hollow square and dashed lines are used to guide the eye, and the thick black line denotes the right boundary, i.e. the middle vertical line in the black-shaded region of Fig.~\ref{fig:edge}(b) or (c). Hence we have proved $S^{(2)} \mapsto \sigma^y_{i} \sigma_{i+1}^y$. Similarly, for $S^{(1)} = \tilde{\tau}^x_i \tilde{\sigma}^z_{i+1} \tilde{\tau}^x_{i+2}$, the map gives us 
\begin{widetext}
    \begin{equation}
        \begin{split}
            S^{(1)} = \tilde{\tau}_i^x \tilde{\sigma}_{i+1}^z \tilde{\tau}_{i+2}^x \;&\equiv \;\; \TikzAppA
            \;\;\leftrightarrow \;\;\TikzAppB \;\; \times\;\; \TikzAppC \;\; \times \;\; \TikzAppD \\
            &\;\:= \TikzAppE \;\;\times \;\; \TikzAppD \;\;=\;\; \TikzAppF = B_s = \prod_{i\in s}  \sigma_{i\in s}^z
        \end{split}
    \end{equation}
\end{widetext}
where we have used the solid green square and the solid red square to denote $\tilde{\tau}^x$ and $\tilde{\sigma}^z$ of $\tilde{\Lambda}$; and used the solid green, red, and blue bullets to denote $\sigma^y$, $\sigma^x$, and $\sigma^z$ of $\Lambda$ respectively. 
The hollow green/red squares and dashed lines are used to guide the eye, and the thick black line denotes the left boundary of the right-half of the torus, i.e. the middle vertical line in the black-shaded region of Fig.~\ref{fig:edge}(b) or (c). Hence we have proved $S^{(1)} \mapsto B_s$ as claimed in Sec.~\ref{sec:edge}.

\bibliography{biblio.bib}
\end{document}